\documentclass[letterpaper,twoside,10pt]{article}
\usepackage{amssymb}
\usepackage{amsmath}
\usepackage{latexsym}
\usepackage{epsfig}
\usepackage{fancyheadings}
\usepackage{pstricks,pst-node}
\usepackage{rotating,graphicx}
\newlength{\figurewidth}
\newlength{\enviropost}
\newcommand{\be}{\begin{equation}}
\newcommand{\ee}{\end{equation}}
\newcommand{\ble}[1]{\begin{equation} \label{#1}}
\newcommand{\bae}{\begin{eqnarray}}
\newcommand{\eae}{\end{eqnarray}}
\newcommand{\fle}[2]%
{\vspace{1.5ex}
\be
\label{#1}
\mbox{%
\setlength{\fboxsep}{3ex}%
\framebox{$\dss #2 $}}
\ee} 

\newcommand{\nn}{\nonumber}
\newcommand{\ff}{\nn \\}
\newcommand{\fe}{& = &}

\newtheorem{state}{S$\! \!$}
\newtheorem{defin}{D$\! \!$}
\newtheorem{exatitle}{Example}
{\noindent \textsc{Proof}:\ }%
{\hfill $\blacksquare$  \vspace{\enviropost} \\}
{\begin{exatitle} \label{#2} #1 \end{exatitle}}%
{\hfill $\Box$ \vspace{\enviropost} \\}
{\vspace{3mm} \noindent {\bf Aside:} \small}%
{\hfill \rule[-3mm]{0mm}{0mm}$\Diamond$\\}
{%
\vspace{3mm}  
\begin{center}
\begin{minipage}{.8\textwidth} 
\begin{state} 
\label{#1}%
}%
{%
\end{state}
\end{minipage}
\end{center}
\vspace{3mm}
}
{%
\vspace{3mm}  
\begin{center}
\begin{minipage}{.8\textwidth} 
\begin{defin} 
\label{#1}%
}%
{%
\end{defin}
\end{minipage}
\end{center}
\vspace{3mm}
}

\newcommand{\dss}{\displaystyle}

\newcommand{\ket}[1]{| #1 \rangle}
\newcommand{\bra}[1]{\langle #1 |}
\newcommand{\ipq}[2]{\left\langle #1 \, | \, #2\right\rangle}

\newcommand{\rt}{\sqrt{2}}

\newcommand{\eg}{\hbox{\em e.g.{}}}

\newcommand{\ie}{\hbox{\em i.e.{}}}
\newcommand{\wrt}{\hbox{w.r.t.{}}}

\newcommand{\rhs}{\hbox{r.h.s.{}}}

\newcommand{\capitem}[1]{\caption{\textsf{#1}}}

\newcommand{\papertitle}{%
Spin 1/2 Particle on a Cylinder with Radial Magnetic Field%
}
\newcommand{\paperauthor}{%
C.{} Chryssomalakos, 
A.{} Franco, and A.{} Reyes-Coronado%
}
\begin{document}
\begin{titlepage}
\vspace*{-1cm}
\begin{flushright}
\textsf{}
\\
\textsf{}
\\
\mbox{}
\\
\textsf{July 8, 2003}
\\[22mm]
\end{flushright}
\renewcommand{\thefootnote}{\fnsymbol{footnote}}
\begin{LARGE}
\bfseries{\sffamily \papertitle}
\end{LARGE}

\noindent \rule{\textwidth}{.6mm}

\vspace*{1.6cm}

\noindent \begin{large}
\textsf{\bfseries%
Chryssomalis Chryssomalakos,
}
\end{large}


\phantom{XX}
\begin{minipage}{.8\textwidth}
\begin{it}
\noindent Instituto de Ciencias Nucleares\\
Universidad Nacional Aut\'onoma de M\'exico\\
Apdo.{} Postal 70-543, 04510 M\'exico, D.F., MEXICO
\vspace{1mm}\\
\end{it}
\texttt{%
chryss@nuclecu.unam.mx%
} 
\phantom{X}
\end{minipage}
\\

\vspace*{15mm}

\noindent \begin{large}%
\textsf{\bfseries%
Alfredo Franco, and Alejandro Reyes-Coronado
}
\end{large}


\phantom{XX}
\begin{minipage}{.8\textwidth}
\begin{it}
\noindent Instituto de F\'{\i}sica\\
Universidad Nacional Aut\'onoma de M\'exico\\
Apdo.{} Postal 20-364, 01000 M\'exico, D.F., MEXICO
\vspace{1mm}\\
\end{it}
\texttt{%
alfredof@fisica.unam.mx,
coronado@fisica.unam.mx%
} 
\phantom{X}
\end{minipage}
\\

\vspace*{2cm}
\noindent
\textsc{\large Abstract: }
We study the motion of a quantum charged particle, constrained
on the surface of a cylinder, in the presence of a radial
magnetic field. When the spin of the particle is neglected,
the system essentially reduces to an
infinite family of simple harmonic oscillators, equally spaced
along the axis of the cylinder. 
Interestingly enough, it can be used
as a quantum Fourier transformer, with convenient visual
output. When the spin 1/2 of the
particle is taken into account, a non-conventional perturbative 
analysis results in a recursive closed form for the
corrections to the energy and the wavefunction, for all
eigenstates, to all orders
in the magnetic moment of the particle. A simple
two-state system is also presented, the time evolution of
which involves an approximate precession of the spin 
perpendicularly to the magnetic field. A number of
plots highlight the findings while several
three-dimensional animations have been made available on the web. 
\end{titlepage}
\setcounter{footnote}{0}
\renewcommand{\thefootnote}{\arabic{footnote}}
\setcounter{page}{2}
\noindent \rule{\textwidth}{.5mm}

\tableofcontents

\noindent \rule{\textwidth}{.5mm}
\section{Introduction}
\label{Intro}
The quantum mechanical description of the motion
of charged particles in a magnetic field is a
classic application of the
theory, having given rise to some of its most
striking results.
Among them, the seminal analysis by
Dirac~\cite{Dir:31},
of the motion in
the field of a magnetic monopole, continues to
inspire
decades after its inception, and
motivates the study of similar quantum systems
that share the
characteristic
of providing insights into the fundamentals
without too much distraction by analytical
complexity.
Such systems are
invaluable pedagogically, as they furnish a
manageable, yet captivating testing ground of
the fundamentals of the theory.

The problem of the motion of a non-relativistic
quantum particle in a plane,
in the presence of a perpendicular homogeneous
magnetic field
is presented in several textbooks (see,
\eg,~\cite{Lan.Lif:81}) --- nevertheless, it
seems to be the
only standard example of this type available.
The main purpose of
this paper is to draw attention to the fact that
the
analogous problem for the cylinder is also
manageable, even
when augmented to include a spin 1/2. In this
latter case, we
also show how the use of the creation and
anihilation operator
machinery greatly simplifies the perturbative
analysis of
the problem, in comparison to the standard
textbook
procedure.

Despite the simplicity of the problem and it
being an obvious variation on the monopole
theme, we have
not been able to find a treatment in the
literature. The motion of a spin-1/2 particle in
the field
of a magnetic monopole has been studied in
detail, both in the
non-relativistic
\cite{Ban:46,For.Whe:59,Gol:65,Sch.Mil.Tsa.DeR.Cla:76,Har:48}
and
relativistic~\cite{Kaz.Yan.Gol:77} cases.
Symmetry aspects of the
problem have also been considered extensively
(see,
\eg,~\cite{Jac:80}), with the discovery of an
underlying
supersymmetry among the most notable
results~\cite{Hok.Vin:84,Cro.Rit:83,Hay.Rau:85}.
On the other hand, quantum {\em spinless} particles moving on
curves or surfaces have been extensively studied
(see, \eg, ~\cite{Cos:81,Tak.Tan:92,Sch.Jaf:03}
and references therein) with a general discussion of the
effects of a vector potential given in~\cite{Enc.Nea:99}. 
It is our hope that the
use of the above simple system will enhance the
exposition of this fascinating part of the
theory. It should also be of interest in
practical applications, such as constrained
quantum mechanics and carbon nanotube physics. 

Consider a classical charged particle,
constrained to move on the
surface of an infinite cylinder, in the presence
of a radial
magnetic field,
\ble{Bfield}
\vec{B}(\vec{r}) = B_0 \frac{a}{\rho} \hat{\rho}
\, ,
\ee
where $a$ is the radius of the cylinder and
$B_0$ is the
field strength on its surface%
\footnote{%
Such a radial field can be thought to be
produced by a homogeneous
linear magnetic charge density, or,
more realistically, in the exterior of a
solenoid of radius $R$,
$R < a$, placed along the
axis of the cylinder and carrying surface
current density $\vec{J}=-
\frac{2 B_0 a}{\mu_0 R^2} z \hat{\phi}$.%
}.
The equations of motion for the particle are
\ble{eomc}
m \dot{v}_z = -qB_0 \, v_\phi
\, ,
\qquad
\qquad
m \dot{v}_\phi = qB_0  v_z
\, ,
\ee
where $m$, $q$ are the mass and charge of the
particle respectively
and $\{ \hat{\rho}, \hat{\phi}, \hat{z} \}$ is a
right-handed
basis. The solutions to (\ref{eomc}) are two
simultaneous oscillations:
the momentum $p_z$ of the particle oscillates
like,
say, $\cos(\omega
t)$ (with $\omega = qB_0/mc$) while its angular
momentum along the
$z$-axis oscillates like $\sin(\omega t)$.
Thus, the particle's kinetic
energy oscillates between a linear and a
rotational
form, becoming, for example,
purely rotational at the turning points of the
oscillation along $z$.

We study, in this
paper, the quantum mechanical version of the
above
problem, adding, at a later stage, a spin-1/2 to
the particle.
The treatment of the spinless case, contained in
Sect.~\ref{TSC},
is exact --- the problem separates
and reduces to an infinite collection of
harmonic
oscillators along $z$.
We find, nevertheless, the resulting quantum
system particularly
rich and with surprising properties ---
it functions, for example, as a
quantum Fourier transformer with convenient
visual output (see
Sec.~\ref{TqFt}). The addition
of spin is treated perturbatively in
Sect.~\ref{TSOHC}, with a
non-conventional method that greatly simplifies
the calculations.
We are able to
give recursion relations for the corrections to
the wavefunctions and
the energy to all orders, for all unperturbed
eigenstates, and apply
the results to compute second-order corrections
to the ground state.
Several plots highlight the findings. We also
make available
on the web several three-dimensional color
animations of the time
evolution of the wavefunction, with or without
spin, and
corresponding to various
initial conditions. An appendix shows how the
standard perturbation theory treatment of the
problem
reproduces, albeit laboriously, our first order
results.
\section{The Spinless Case}
\label{TSC}
\subsection{The spectrum}
\label{Ts}
The magnetic field of Eq.~(\ref{Bfield}) can be obtained, in the
vicinity of the surface of the cylinder, from the
vector potential
\ble{Afield}
\vec{A}(\vec{r}) = -B_0 \frac{a}{\rho} z \hat{\phi}
\, .
\ee
The Hamiltonian for a quantum spinless 
particle constrained to move on
the surface of the cylinder is given by
\bae
\label{Hamiltonian}
\hat{H} 
\fe
\frac{1}{2m}(\vec{p} - \frac{q}{c} \vec{A})^2
\ff
 \fe 
- \frac{\hbar^2}{2m} (\frac{1}{a^2} \partial_\phi^2 + \partial_z^2) 
+ \frac{q^2 B_0^2}{2mc^2} \, z^2 
- i \frac{\hbar q B_0}{mca} z \partial_\phi
\, . 
\eae
The wavefunction $\Psi(\phi,z)=\frac{1}{\sqrt{2 \pi}} 
e^{i \ell \phi} Z(z)$ is an
eigenfunction of $\hat{H}$, with eigenvalue $E$, provided  
$Z(z)$ satisfies (primes denote differentiation \wrt{} $z$)
\ble{Zeq}
- \frac{\hbar^2}{2m} Z''(z) 
+ \frac{1}{2} m \omega^2 (z+\ell b)^2 Z(z) 
= E Z(z)
\, ,
\ee
where we have set
\ble{ombdef}
\omega = \frac{q B_0}{mc}
\, ,
\qquad
\qquad
\qquad
b = \frac{\hbar c}{q B_0 a}
\, ,
\ee
and, in what follows, we take $\hbar=m=\omega=1$.
This is the equation for a simple harmonic oscillator (SHO), centered 
at $z=-\ell b$. We conclude that, for each integer value 
of $\ell$, one
obtains a copy of the usual SHO spectrum, centered at $z=-\ell b$,
\ie, the eigenfunctions and eigenvalues of $\hat{H}$ are given by 
\ble{eigenfun}
\ipq{\phi, \,z}{n,\, \ell}=
N_n H_n(z+\ell b) e^{-(z+\ell b)^2/2} \frac{1}{\sqrt{2 \pi}} 
e^{i \ell \phi}
\, ,
\qquad
\qquad
E_{\ell, n} \equiv E_n = n+\frac{1}{2}
\, ,
\qquad
\qquad
N_n \equiv (2^n n! \sqrt{\pi})^{-\frac{1}{2}}
\, ,
\ee
where $\ket{n,\, \ell}$ denote the corresponding
eigenkets ($n=0,1,2,\dots$; $\ell \in \mathbb{Z}$) 
and $H_n(z)$ are the Hermite polynomials. 
\subsection{A quantum Fourier transformer}
\label{TqFt}
Suppose that the wavefunction of the particle, at $t=0$, is given by%
\footnote{%
We use the notation 
$\ipq{\phi, \,z}{n,\, \ell,\, m}
=
N_n  H_n(z+\ell b) e^{-(z+\ell b)^2/2}
\frac{1}{\sqrt{2 \pi}} e^{i m \phi}$ (not to be confused,
hopefully, with the
standard spherical symmetry notation)--- these wavefunctions are 
eigenfunctions of $\hat{H}$
only when $m=\ell$, in which case they will be denoted by $\ket{n,\,
\ell}$, as above.%
}
\ble{wft0}
\Psi_\ell(\phi, \, z, \, t=0) 
\equiv 
\ipq{\phi, \, z}{0,\, 0, \, \ell}
= 
N_0 e^{-z^2/2} \frac{1}{\sqrt{2 \pi}} 
e^{i \ell \phi}
\, .
\ee
If $\ell =0$, we have one of the infinitely many ground states of the
system and the time evolution is by a phase factor. Consider now the
case $\ell\neq 0$. Then the $z$-part ``sees'' a quadratic potential
centered at $z=-\ell b$ but the initial wavefunction is a gaussian
centered at the origin. This is a coherent state and its time
evolution is an oscillation around $z=-\ell b$, with the frequency
$\omega=1$ of the oscillator,
\ble{timeev}
\Psi_\ell(\phi, \, z, \, t)= 
N_0 e^{-it/2} 
\, e^{-i (z+\ell b) \, \ell b \sin t} 
\, e^{-(z + \ell b (1- \cos t))^2/2}
\, \frac{1}{\sqrt{2 \pi}} e^{i \ell \phi}
\, .
\ee 
Notice that a physically irrelevant global phase factor $e^{i
\ell^2 b^2/4 \sin 2t}$ has been omitted from the above
expression. 
We may now exploit linearity to write down the time evolution of a
gaussian (in $z$), centered at the origin,  with arbitrary 
$\phi$-dependence,
\ble{Psiarb}
\Psi(\phi, \, z, \, t=0) =  N_0 e^{-z^2/2} f(\phi)
\, .
\ee
Writing
\ble{fFmodes}
f(\phi) = \sum_{\ell= -\infty}^{\infty} f_\ell e^{i \ell \phi}
\, ,
\ee
we obtain,
\ble{timeevgen}
\Psi(\phi, \, z, \, t)=
\sum_{\ell= -\infty}^{\infty} f_\ell \Psi_\ell(\phi, \, z, \, t)
\, ,
\ee
\ie, each Fourier mode of $f(\phi)$ gives rise to a gaussian in $z$,
oscillating like a coherent state around $z=-\ell b$ with frequency
$\omega=1$. Taking $b \gg 1$, so that the various
gaussians separate after half a period, converts the system to a 
quantum Fourier
transformer with convenient visual output: 
looking at the wavefunction 
at time $t=\pi$ 
(a half-period), one sees the above gaussians at the (second) 
turning point of 
their oscillation, at $z=-2\ell b$, 
with their amplitudes proportional 
to the Fourier amplitudes $f_\ell$. 
\begin{figure}
\includegraphics[width=.97\textwidth]{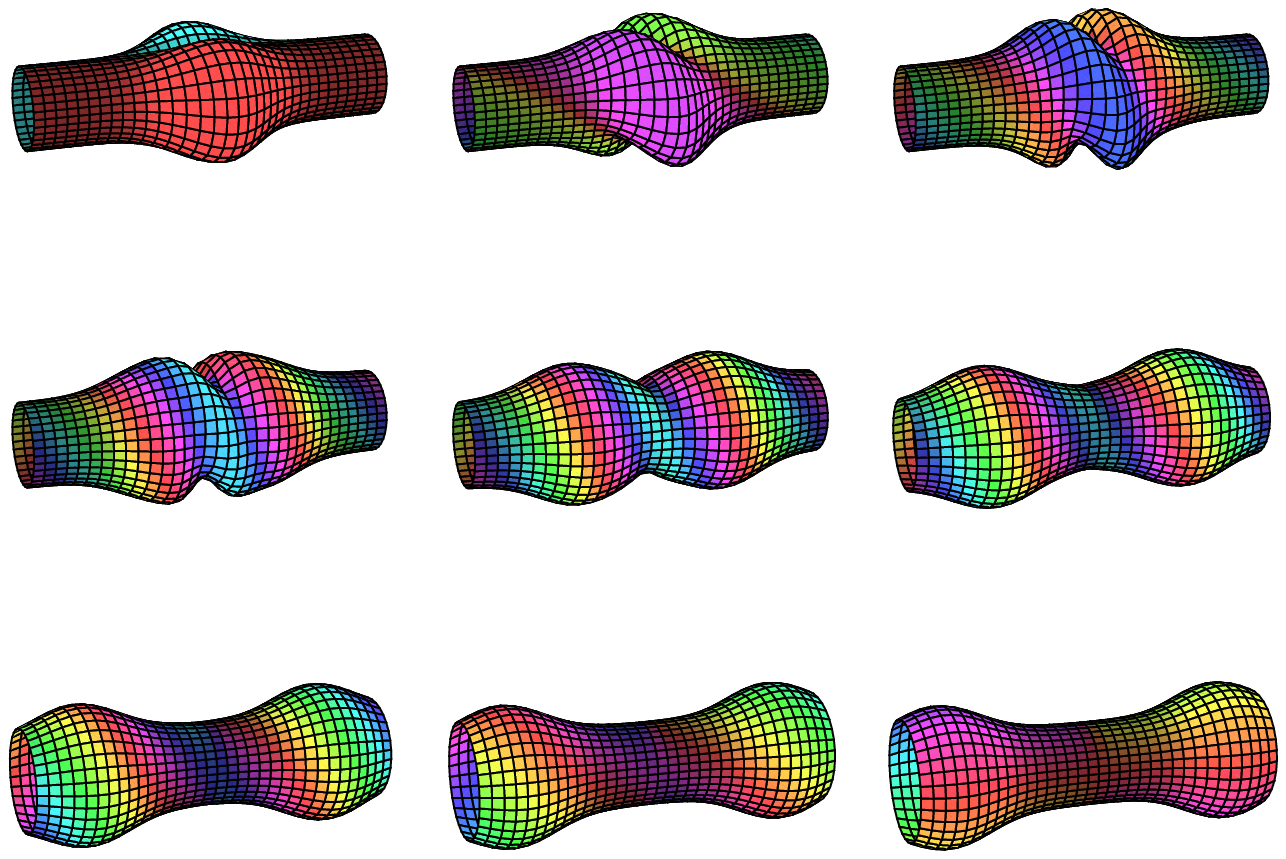}
\capitem{\textbf{Time evolution of the initial wavefunction
\mathversion{bold} 
$\Psi(\phi, \, z, \, t=0) =  N_0 e^{-z^2/2} \cos \phi$. 
\mathversion{normal}}
The modulus
of $\Psi$ is indicated by the radial distance of the surface from that
of the
cylinder while its phase is color-coded, with $1$, $i$, $-1$, $-i$
corresponding to red, green, blue, purple 
(several animations in color, including the
above, can be seen at
\texttt{http://www.nuclecu.unam.mx/$\sim$chryss}). The time $t$
is equal to zero at the top
left and increases to the right and downwards, reaching
$t=\pi$ (half a period) at the bottom right.%
}
\label{osci2}
\end{figure}
In Fig.~\ref{osci2}, we plot several frames of
the time evolution of $\Psi$, when $f(\phi)=\cos \phi$ 
--- the last frame,
at $t=\pi$, clearly displays the Fourier content of $f$. 
Fig.~\ref{osci3} corresponds to the initial
wavefunction 
$\Psi(\phi, \, z, \, t=0) 
\sim
e^{-z^2/2} 
(1 + e^{-i \phi} + 1.5 e^{-i 2 \phi} + e^{-i 3 \phi})$.
\begin{figure}
\includegraphics[width=\textwidth]{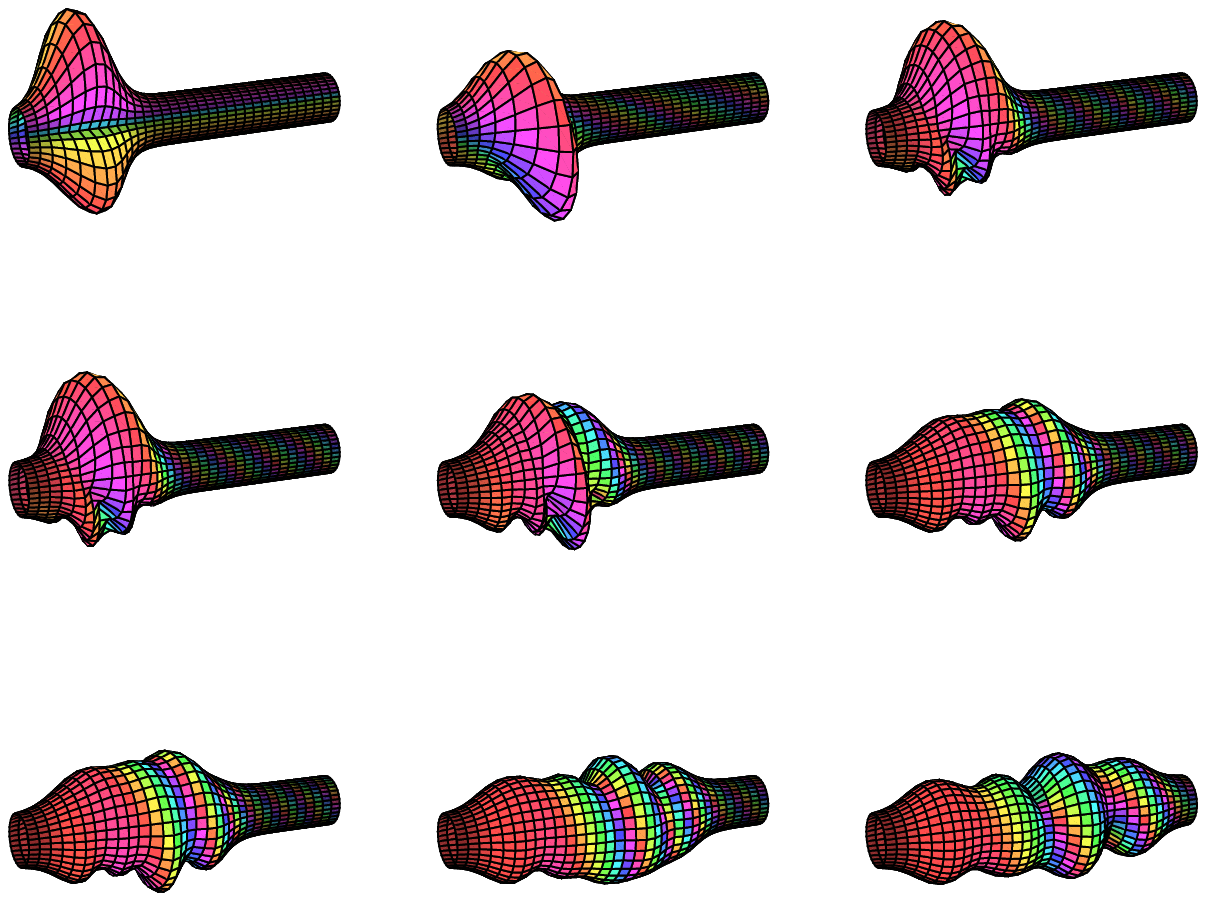}
\capitem{\textbf{Time evolution of the initial wavefunction
\mathversion{bold} 
$\Psi(\phi, \, z, \, t=0)  \sim e^{-z^2/2} 
(1 + e^{-i \phi} + 1.5 e^{-i 2 \phi} + e^{-i 3 \phi})$. 
\mathversion{normal}}
Conventions are as in Fig.~\ref{osci2}. Notice how the
amplitudes of the gaussians in the last frame
correspond to the Fourier components of the initial
wavefunction.%
}
\label{osci3}
\end{figure}
\section{The Spin-1/2 Case}
\label{TSOHC}
\subsection{Separation of variables}
\label{Sov}
For a spin-1/2 particle, the wavefunction has two components,
$\Psi_+(\phi, \, z)$,  $\Psi_-(\phi, \, z)$, which we arrange in a
column vector. The spin interacts with the magnetic field via
$\hat{H}_\text{int} = -\lambda \vec{S} \cdot \vec{B}$, 
which, for the
field given in~(\ref{Bfield}) becomes
\ble{Hint}
\hat{H}_\text{int} = - \frac{1}{2} \lambda B_0 
\left(
\begin{array}{cc}
0 & e^{-i \phi} \\
e^{i \phi} & 0 
\end{array}
\right)
\, .
\ee
In order to achieve separation of variables now, we need to take 
\ble{Psipm}
\Psi_+(\phi,\, z) = \frac{1}{\sqrt{2 \pi}} e^{i \ell \phi} Z_+(z)
\, ,
\qquad
\qquad
\Psi_-(\phi,\, z) = 
\frac{1}{\sqrt{2 \pi}} e^{i (\ell+1) \phi} Z_-(z)
\, .
\ee
The resulting equation for the $Z$'s is
\bae
\label{Zpmeq}
\hat{H}_\ell Z_+(z) + \epsilon Z_-(z) 
\fe
 E Z_+(z)
\ff
\hat{H}_{\ell+1} Z_-(z) + \epsilon Z_+(z)
\fe
E Z_-(z)
\, ,
\eae
where $\hat{H}_\ell$ is a SHO hamiltonian centered at $z=-\ell b$,
\ble{Hell}
\hat{H}_\ell = 
-\frac{1}{2} \partial_z^2 + \frac{1}{2} (z+\ell b)^2
\, ,
\ee
and $\epsilon \equiv -\hbar \lambda B_0/2$.
We see that the problem reduces to that of two SHO's,  
a distance $b$
apart, coupled by the $\epsilon$ terms in~(\ref{Zpmeq}). 
Our task is to
solve~(\ref{Zpmeq}) perturbatively in $\epsilon$. 
Once the solutions are
known, to a certain order in $\epsilon$, we can form the spinor 
\ble{spinor}
\Psi(\phi,\, z) = 
\left(
\begin{array}{c}
\Psi_+(\phi,\, z)       
\\
\Psi_-(\phi,\, z)
\end{array}
\right)
=
\left(
\begin{array}{c}
\frac{1}{\sqrt{2 \pi}} e^{i \ell \phi} Z_+(z)       
\\
\frac{1}{\sqrt{2 \pi}} e^{i (\ell+1) \phi} Z_+(z)
\end{array}
\right)
\, ,
\ee
from which the probability density and spin direction 
can be extracted
as
\ble{densdir}
\rho= \sqrt{|Z_+|^2 + |Z_-|^2}
\, ,
\qquad
\qquad
\alpha = 2 \arctan \frac{|Z_-|}{|Z_+|}
\, ,
\qquad
\qquad
\beta= \text{Im}\, (\log \frac{\Psi_-}{\Psi_+})
\, ,
\ee
where the spin direction
$\hat{n}$ is given by  
$\hat{n} = (\sin \alpha  \cos \beta, \, \sin \alpha \sin \beta, \,
\cos \alpha)$, in Cartesian coordinates. 
The solutions of~(\ref{Zpmeq}) 
have no relative
(complex) phase and can be taken real. Then, on the $\phi=0$ plane,
the spin lies in the $x-z$ plane. 
The extra $e^{i\phi}$ factor in $\Psi_-$
guarantees that when we change our position on the cylinder by
$\phi$, the spin also rotates by the same angle and its direction is
therefore obtained by revolution of the $\phi=0$ configuration, in
other words, $\beta=\phi$ in~(\ref{densdir}). These
remarks are of course valid only for the energy eigenstates --- the
time evolution of general states results in the spin pointing outside
of the radial plane as well, even if they start within it. 
\subsection{Perturbative solution}
\label{Ps}
One may treat the system of the two coupled differential equations
in~(\ref{Zpmeq}) by standard perturbation theory methods --- we give
the first-order analysis along these lines in the appendix. It is
instructive though, as well as much more efficient, 
to exploit the SHO
machinery of raising and lowering operators. 
We begin by transforming~(\ref{Zpmeq}) into a single 
differential-difference
equation. Indeed, it is clear from the symmetry of these equations
that the solutions can be taken to satisfy $Z_-(z)=\pm Z_+(-z-b)$ 
--- we will refer to the two possibilities 
as {\em symmetric} and {\em
antisymmetric} respectively. Taking $\ell=0$ and restricting to the
symmetric case, the first of~(\ref{Zpmeq}) becomes
\ble{Zpeq}
\hat{H}_0 Z_{\text{s}+}(z) + \epsilon Z_{\text{s}+}(-z-b) =
EZ_{\text{s}+}(z)
\, .
\ee
We now write $\hat{H}_0=a^{\dagger} a + 
\frac{1}{2}$ and introduce the
ket whose wavefunction is $Z_{\text{s}+}(z)$,
\ble{Zpket}
Z_{\text{s}+}(z) = \ipq{z}{Z_{\text{s}+}} 
\, ,
\qquad
\ket{Z_{\text{s}+}} = \sum_{n=0}^\infty c_n (a^\dagger)^n \ket{0}_0
\equiv f(a^\dagger) \ket{0}_0
\, ,
\ee
where $\ket{0}_0$ is the ground state for $\ell$ equal to zero (\ie,
centered at the origin) and $f(a^\dagger)$ is defined by the last
equation --- the idea is that any ket can be 
obtained by some function of
$a^\dagger$ applied to the ground state. The wavefunction
$Z_{\text{s}+}(-z-b)$ is obtained from $Z_{\text{s}+}(z)$ by first
reflecting around the origin and then effecting the translation
$z\mapsto z+b$. Reflecting around the origin an eigenfunction of
$\hat{H}_0$ introduces a sign given by the parity of the state, which
shows that the reflected state is produced by $f(-a^\dagger)$ 
applied
to the ground state. The reflected and translated state then is given
by
\bae
\label{reftran}
T_{-b} f(-a^\dagger) \ket{0}_0 
\fe
 e^{-b^2/4} e^{-\frac{b}{\rt}a^\dagger} e^{\frac{b}{\rt} a} 
f(-a^\dagger) \ket{0}_0
\ff
 \fe 
e^{-b^2/4} e^{-\frac{b}{\rt}a^\dagger} f(-a^\dagger
-\frac{b}{\rt})  \ket{0}_0
\, ,
\eae
which brings~(\ref{Zpeq}) into the form
\ble{Zpeq2}
(a^\dagger a+ \frac{1}{2}) f(a^\dagger) \ket{0}_0 
   +  \epsilon e^{-b^2/4}
      e^{-\frac{b}{\rt}a^\dagger} 
      f(-a^\dagger -\frac{b}{\rt})  \ket{0}_0 
= 
E f(a^\dagger) \ket{0}_0
\, .
\ee
Noting that the $a - a^\dagger$ commutation relations are identical
to the $\partial_x - x$ ones, we may infer a differential -
difference equation for the function $f(x)$,
\ble{diffdiff}
x f'(x) 
+ \frac{1}{2} f(x) 
+ \epsilon e^{-b^2/4} e^{-\frac{b}{\rt}x}  f(-x -\frac{b}{\rt})
=
E f (x)
\, .%
\ee
We now specify to the case where the unperturbed state is the $n$-th
excited state of the SHO. The perturbed state will be denoted by
$f_n(a^\dagger)\ket{0}_0$, with energy $E_n$, where
\ble{power}
f_n(x) = \sum_{k=0}^\infty f_n^{(k)}(x) \epsilon^k
\, ,
\qquad
\qquad
\qquad
E_n = \sum_{k=0}^\infty E_n^{(k)} \epsilon^k 
\, .
\ee
Notice that 
\ble{f0E0}
f_n^{(0)}(x)=\frac{1}{\sqrt{n!}} x^n
\, ,
\qquad
\qquad
\qquad
E_n^{(0)} =n+\frac{1}{2}
\, .
\ee
Substituting these expansions in~(\ref{diffdiff}) we obtain
\bae
\label{fnk}
x\partial_x f_n^{(k)}(x) -nf_n^{(k)}(x)
\fe
{}-e^{-b^2/4} e^{-\frac{b}{\rt}x} f_n^{(k-1)}(-x-\frac{b}{\rt})
+\frac{1}{\sqrt{n!}} E_n^{(k)} x^n 
\ff
 & &
{}+ \sum_{m=1}^{k-1} E_n^{(k-m)} f_n^{(m)}(x)
\, ,
\eae
where we separated the $m=0$, $k$ terms in the sum on the \rhs{} and
used~(\ref{f0E0}). Notice that the \rhs{} above only contains
$f_n^{(m)}$ with $m<k$,  so~(\ref{fnk}) can be used recursively
to determine any $f_n^{(k)}(x)$. 

The requirement that the perturbed eigenket be
normalized implies that the corrections, order by order in $\epsilon$,
have to be orthogonal to the unperturbed eigenket $\ket{n}_0$. This in
turn implies that, for $k > 0$, the coefficient of $x^n$ in
$f_n^{(k)}(x)$ must vanish. 
Then so does the coefficient of $x^n$ in $x
\partial_x f_n^{(k)}(x)$. Using this information, we can extract the
coefficient of $x^n$ on both sides of~(\ref{fnk}) --- the resulting
equation fixes recursively the energy corrections $E_n^{(k)}$,
\ble{Enk}
E_n^{(k)} 
= 
\frac{1}{\sqrt{n!}} e^{-b^2/4} 
\sum_{r=0}^{n} \binom{n}{r} \big( -\frac{b}{\rt} \big)^{n-r} 
(\partial_x^r f_n^{(k-1)})(-\frac{b}{\rt})
\, .
\ee
This is an appropriate point to comment on the antisymmetric
solutions. The difference in this case is that the $\epsilon$ term
in~(\ref{diffdiff}) appears with a minus sign. Since this is the only
place where $\epsilon$ appears explicitly, we conclude that one gets
the antisymmetric solutions from the symmetric ones by the
substitution $\epsilon \rightarrow -\epsilon$. As we will see later
on, symmetric solutions have their spin parallel, more or less, with
the magnetic field while antisymmetric ones have it antiparallel.
Since $\epsilon$ is proportional to the magnetic moment of the
particle, the above statement about the relation between the two kinds
of solutions essentially says that the symmetric solution for a
particle coincides with the antisymmetric solution for the same
particle but with the opposite magnetic moment. 
\subsection{Corrections to the ground state}
\label{Cttgs}
For $n=0$, Eqs.~(\ref{fnk}), (\ref{Enk}) simplify considerably,
\bae
x \partial_x f_0^{(k)}(x)  
\fe
{}-e^{-b^2/4} e^{-\frac{b}{\rt}x} f_0^{(k-1)}(-x-\frac{b}{\rt})
+ E_0^{(k)}  
+ \sum_{m=1}^{k-1} E_0^{(k-m)} f_0^{(m)}(x)
\label{f0k}
\\
E_0^{(k)} 
\fe 
e^{-b^2/4} f_0^{(k-1)}(-\frac{b}{\rt})
\label{E0k}
\, .
\eae
The unitarity argument given above implies in this case that
$f_0^{(k)}(0)=0$, for all $k$ greater than zero --- this fixes the
lower integration limit in the solution of~(\ref{f0k}) equal to zero. 
The change of
variable $x'_k \rightarrow x s_k$ and the substitution
of~(\ref{E0k}) finally give 
\ble{f0kf}
f_0^{(k)}(x) = 
e^{-b^2/4} \int_0^1 \frac{ds_k}{s_k} \left\{
f_0^{(k-1)}(-\frac{b}{\rt}) 
- e^{-\frac{b}{\rt} x s_k} 
f_0^{(k-1)}(-x s_k -\frac{b}{\rt})
+ \sum_{m=1}^{k-1} f_0^{(k-m-1)}(-\frac{b}{\rt}) f_0^{(m)}(x s_k)
\right\} 
\, .
\ee
Notice that the (apparent) pole 
of the integrand at $s_k=0$ cancels out.

We look in some
detail now at the wavefunctions and resulting spin configurations,
including up to quadratic corrections. For the first three
$f^{(k)}$, Eqs.~(\ref{f0E0}), (\ref{f0kf}) give
\bae
f_0^{(0)}(x)
\fe
1
\label{f00}
\\
f_0^{(1)}(x)
\fe
e^{-b^2/4} \int_0^1 
\frac{ds_1}{s_1} \left( 1-e^{-\frac{b}{\rt} x s_1}
\right)
\label{f01}
\\
f_0^{(2)}(x)
\fe
e^{-b^2/2} \int_0^1 \int_0^1
\frac{ds_2 ds_1}{s_2 s_1} 
\left\{ 2 -e^{b^2 s_1/2} -e^{-\frac{b}{\rt}
x s_2} + e^{\frac{b^2}{2} s_1} e^{-\frac{b}{\rt} s_2(1-s_1)x}
-e^{-\frac{b}{\rt}s_2 s_1 x} \right\}
\, ,
\label{f02}
\eae
while for the corresponding energy corrections we get
\ble{E0012}
E_0^{(0)}=\frac{1}{2}
\, ,
\qquad
\qquad
E_0^{(1)} =e^{-b^2/4}
\, ,
\qquad
\qquad
E_0^{(2)} = 
e^{-b^2/2} \int_0^1 \frac{ds_1}{s_1} 
\left( 1-e^{\frac{b^2}{2} s_1} \right)
\, .
\ee
Applying the above $f$'s to the ground state and projecting onto the
position eigenket $\ket{z}$ we find the wavefunctions 
\bae
Z_+^{(0)}(z)
\fe
N_0 
e^{-z^2/2}
\label{Z0Sol}
\\
Z_+^{(1)}(z)
\fe
N_0 
e^{-b^2/4} 
\int_0^1 \frac{ds_1}{s_1} 
\left( 
e^{-z^2/2}
-e^{b^2 s_1^2/4} e^{-(z+b s_1)^2/2}
\right)
\label{Z1Sol}
\\
Z_+^{(2)}(z)
\fe
N_0
e^{-b^2/2} 
\int_0^1 \int_0^1 \frac{ds_2 ds_1}{s_2 s_1} 
\left\{ 
\left(
2 -e^{b^2/2 s_1} 
\right)
e^{-z^2/2}
-e^{b^2s_2^2/4} e^{-(z+bs_2)^2/2}
\right.
\ff
 & & 
\left. {}+ e^{b^2 s_1/2 
+b^2s_2^2(1-s_1)^2/4} e^{-(z+bs_2(1-s_1))^2/2}
-e^{b^2s_2^2 s_1^2/4} e^{-(z+bs_2s_1)^2/2}
\right\}
\label{Z2Sol}
\, .
\eae
In terms of these, the symmetric solution for the spinor has
components (we ommit an overall normalization factor)
\ble{spinS}
Z_{\text{s}+}(z) 
= 
Z_+^{(0)}(z) 
+\epsilon Z_+^{(1)}(z) 
+ \epsilon^2  Z_+^{(2)}(z)
\, ,
\qquad
\qquad
Z_{\text{s}-}(z)=Z_{\text{s}+}(-z-b)
\ee
and energy
\ble{ES}
E_{\text{s}0}=E_0^{(0)}+\epsilon E_0^{(1)} +\epsilon^2 E_0^{(2)}
\, ,
\ee
while the antisymmetric solution is given by
\ble{spinA}
Z_{\text{a}+}(z) 
= 
- Z_+^{(0)}(z) 
+ \epsilon Z_+^{(1)}(z) 
- \epsilon^2  Z_+^{(2)}(z)
\, ,
\qquad
\qquad
Z_{\text{a}-}(z)=-Z_{\text{a}+}(-z-b)
\ee
with energy
\ble{EA}
E_{\text{a}0}=E_0^{(0)}-\epsilon E_0^{(1)} +\epsilon^2 E_0^{(2)}
\, .
\ee
Plots of $Z_\pm$, for both cases, as well as the corresponding spin
configurations, are given in Figs.~\ref{ZSspin}, 
\ref{ZAspin}. Notice that the effect of the perturbation is
small, despite a rather large value of $\epsilon$. This can be
traced to the fact that the parameter values used give rise to a
small overlap of two neighboring gaussians. As a result, the
range of validity of our perturbative results is considerable
larger that the standard $\epsilon \ll 1$.
\setlength{\figurewidth}{.97\textwidth}
\begin{figure}
\rule{0mm}{.25\figurewidth}
\begin{pspicture}(0\figurewidth,0\figurewidth)%
                 (.97\figurewidth,.25\figurewidth)
\setlength{\unitlength}{.25\figurewidth}
\psset{xunit=.25\figurewidth,yunit=.25\figurewidth,arrowsize=1.5pt 3}
\put(2.49,0){\makebox[0cm][c]{$z=0$}}
\put(1.51,0){\makebox[0cm][c]{$z=-b$}}
\put(4,0.18){\makebox[0cm][l]{$z$}}
\psline[linewidth=.3mm]{->}%
(0,.257)(4.07,0.257)
\psline[linewidth=.3mm]{->}%
(2.49,0.07)(2.49,0.95)
\psline[linewidth=.3mm,linestyle=dashed]{-}%
(1.51,0.07)(1.51,0.95)
\includegraphics[width=.97\textwidth]{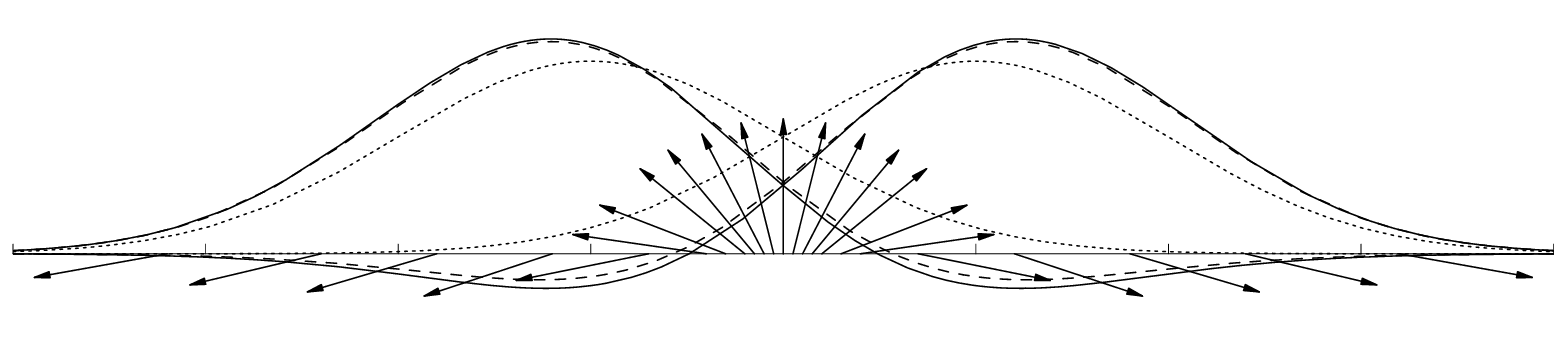}
\end{pspicture}%
\capitem{%
\textbf{The spin components 
\mathversion{bold}
$Z_{\text{s}+}$, $Z_{\text{s}-}$
\mathversion{normal} 
and the
corresponding spin configuration.} The dotted curves give the
zeroth-order result, \ie, two gaussians, centered at $z=0$ and
$z=-b=-2$. The dashed curves include corrections up to first order
while the solid ones up to second ($\epsilon=0.5$ --- 
a rather large value
was used to make the effect visible). The integrals
in Eqs.~(\ref{Z1Sol}), (\ref{Z2Sol}) have been evaluated numerically.
The spin configuration shown includes quadratic corrections. 
Due to the 
symmetry, the spin
points always along $\hat{\rho}$ 
(upwards in the figure) at $z=-b/2=-1$, 
while it tends to $\pm
\hat{z}$ as $z$ tends to $\pm \infty$. 
The perturbation causes a zero 
in each component  (for finite $z$) --- the spin crosses the $z$ 
axis at those points. As a result, the two gaussians are pushed apart
while the width of the central region, where the spin points up, is
reduced.%
}
\label{ZSspin}
\end{figure}
\setlength{\figurewidth}{.97\textwidth}
\begin{figure}
\rule{0mm}{.3\figurewidth}
\begin{pspicture}(0\figurewidth,0\figurewidth)%
                 (.97\figurewidth,.3\figurewidth)
\setlength{\unitlength}{.25\figurewidth}
\psset{xunit=.25\figurewidth,yunit=.25\figurewidth,arrowsize=1.5pt 3}
\put(2.49,0){\makebox[0cm][c]{$z=0$}}
\put(1.51,0){\makebox[0cm][c]{$z=-b$}}
\put(4.02,0.56){\makebox[0cm][l]{$z$}}
\psline[linewidth=.3mm]{->}%
(0,.652)(4.07,0.652)
\psline[linewidth=.3mm]{->}%
(2.49,0.07)(2.49,1.2)
\psline[linewidth=.3mm,linestyle=dashed]{-}%
(1.509,0.07)(1.509,1.2)
\includegraphics[width=.97\textwidth]{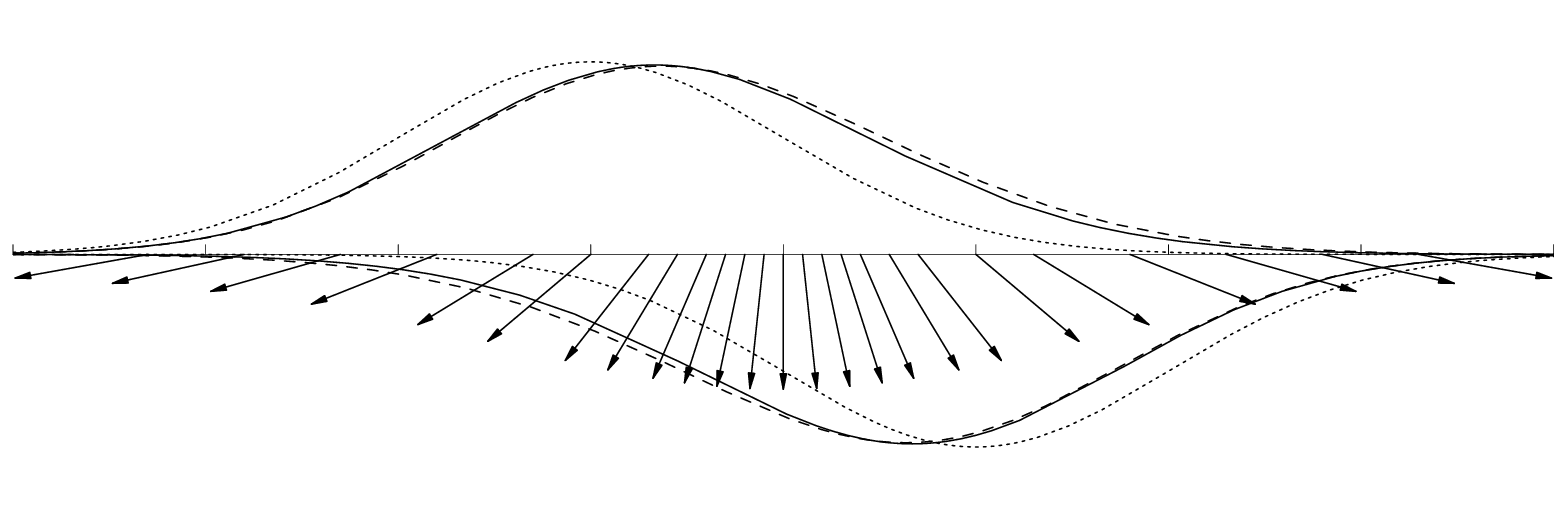}
\end{pspicture}
\capitem{%
\textbf{%
The spin components 
\mathversion{bold}
$Z_{\text{a}+}$, $Z_{\text{a}-}$
\mathversion{normal}
and the corresponding spin configuration.%
} 
Notation and parameter values are as in Fig.~\ref{ZSspin}.
Due to the antisymmetry, the spin
points always along $-\hat{\rho}$ (downwards in the figure) 
at $z=-b/2=-1$, 
while it tends to $\pm \hat{z}$ as $z$ tends to $\pm \infty$. 
The perturbation  pushes  the two gaussians together
while the width of the region where the spin points downwards is
increased.%
}
\label{ZAspin}
\end{figure}
\subsection{A two-state system}
\label{atss}
Consider a state that, at $t=0$, is the sum of the symmetric
and antisymmetric $n=0$ states found above, for, say, $\ell=0$.
The spin-down component of this state is of order 1 while the
spin-up component is of order $\epsilon$. Roughly speaking,
the particle is localized at $z=-b$ and its spin points
along the negative $z$-axis. Then the standard two-state
system analysis shows that the amplitudes to be in the spin-up and
spin-down states at later times behave like $-i \sin(\Omega
t)$ and $\cos(\Omega t)$ respectively, where $\Omega$ 's
expansion in powers of $\epsilon$ starts with $\epsilon
E_0^{(1)}/2$. The spin precesses in the tangent plane to the
cylinder (\ie, perpendicularly to the magnetic field) while
the particle oscillates from $-b$ to zero and back.   
\section{Concluding Remarks}
\label{CR}
We have studied the problem of the motion of a spin-1/2
particle on a cylinder, in the presence of a radial magnetic
field. A non-standard perturbative analysis, applicable to
any perturbation of the harmonic ascillator, led to a recursion
relation for the wavefunction and energy corrections,
Eqs.~(\ref{fnk}) and~(\ref{Enk}) respectively, with explicit
results for the ground state in Eqs.~(\ref{E0012}) --
(\ref{Z2Sol}) and Figures~\ref{ZSspin}, \ref{ZAspin}. 
It is
worth emphasizing that the radial magnetic field of the
problem can be approximated in the laboratory, as pointed out
in the first footnote. 

We end with a comment on 
the form of the unperturbed
hamiltonian used, Eq.~(\ref{Hamiltonian}). When dealing with
the motion of a quantum particle on a surface, one can use a 3-D
Laplacian in the hamiltonian and constrain
the motion of the particle on the surface using a steep
confining potential in the radial direction. It is well known
that, in this approach, which seems to be the one appropriate for
practical applications, there is an induced potential for the motion
along the surface, proportional to the square of the
difference between the two principal curvatures of the
surface (see, \eg,~\cite{Cos:81,Sch.Jaf:03} and
references therein). In our case, this is a constant which only shifts
the energy eigenvalues. Nevertheless, an obvious extension of
our problem here would be the study of the motion on the surface of a
slightly curved cylinder, in which case the above mentioned
induced potential would have to be taken into account. 
\appendix
\section{First Order Corrections to the
Ground State: the Standard Treatment}
\label{SFOPA}
We outline here the standard first order perturbative analysis of
the problem, deriving the corrections to the ground
state. Given that the zeroth order spectrum is
degenerate, we need to first diagonalize the
interaction hamiltonian in each degenerate subspace.
One easily sees that the interaction only connects the
pairs of eigenstates $\ket{n, \, \ell, \, +}$,   
$\ket{n, \, \ell + 1, \, -}$. The appropriate
zeroth order basis is given by the {\em symmetric} and
{\em antisymmetric} linear combinations
\bae
\label{santi}
\ket{n_\ell^\text{s}} 
\, \equiv \, 
\ket{n_\ell^\text{0}} 
\fe 
\frac{1}{\sqrt{2}}
(\ket{n, \, \ell, \, +} + \ket{n, \, \ell + 1, \, -}) 
\ff
\ket{n_\ell^\text{a}} 
\, \equiv \,
\ket{n_\ell^\text{1}} 
\fe
\frac{1}{\sqrt{2}}
(-\ket{n, \, \ell, \, +} + \ket{n, \, \ell + 1, \, -}) 
\, .
\eae
Notice that we use the label $\ell$ for states that
are equally localized at $-\ell b$ and 
$-(\ell + 1) b$. The reason for renaming the states with
numerical superscripts, instead of letters, 
will become apparent below. 
The expectation value of $H_\text{int}$ in these states 
reproduces our
result~(\ref{E0012}) for the first order correction to
the energy. It is interesting to see how the first
order correction to the wavefunction, conventionally
given by an infinite sum, is brought into the
closed form (\ref{Z1Sol}). The matrix elements of
$H_\text{int}$ in the above basis are
\bae
\label{Hintme}
\bra{n_\ell^0} H_\text{int} \ket{m_{\ell'}^0}
\fe
\epsilon \ipq{n_\ell}{m_{\ell + 1}} 
\delta_{\ell \ell'} \delta_{\tilde{n} \tilde{m}} 
\ff
\bra{n_\ell^0} H_\text{int} \ket{m_{\ell'}^1}
\fe
\epsilon \ipq{n_\ell}{m_{\ell + 1}} 
\delta_{\ell \ell'} \delta_{\tilde{n}, \tilde{m}+1} 
\ff
\bra{n_\ell^1} H_\text{int} \ket{m_{\ell'}^1}
\fe
- \epsilon  \ipq{n_\ell}{m_{\ell + 1}} 
\delta_{\ell \ell'} \delta_{\tilde{n} \tilde{m}} 
\, ,
\eae
where $\ipq{n_\ell}{m_{\ell + 1}}$ is the overlap between
SHO eigenstates $\ket{n}$, $\ket{m}$, at a distance $b$ apart 
and $\tilde{n}$ is the parity of $n$.
Specifying to the symmetric ground state and taking $\ell=0$, we
find the first order correction
\ble{focsgs}
\ket{0_0^0}^{(1)}
=
\sum_{k=1}^\infty
\frac{\ipq{0_0}{k_{1}}}{k}
\ket{k_0^{\tilde{k}}}
\, .
\ee
We see that only states with $\ell=0$ contribute.
Furthermore, when $k$ is even, only the symmetric state
contributes while for $k$ odd, only the antisymmetric one does. 
Using the fact that
\ble{zmover}
\ipq{0_0}{k_1} =
\frac{1}{\sqrt{k!}}(-\frac{b}{\sqrt{2}})^k e^{-\frac{b^2}{4}}
\,,
\ee
we find for the spin-up component of the correction
\ble{sufo}
\ket{0_0^0, \, +}^{(1)}
=
\sum_{k=1}^\infty
\frac{(-1)^k}{k \sqrt{k!}} \left (\frac{b}{\sqrt{2}} \right)^k
e^{-\frac{b^2}{4}} \ket{k}
\, ,
\ee
which implies 
\bae
\label{sut}
b \frac{\partial}{\partial b} 
\left(
e^{\frac{b^2}{4}} \ket{0_0^0, \, +}^{(1)}
\right)
\fe
\sum_{k=1}^{\infty} \frac{(-1)^k}{\sqrt{k!}} 
\left( 
\frac{b}{\sqrt{2}} 
\right)^k
\ket{k}
\ff
 \fe
e^{\frac{b^2}{4}} \left( \ket{0_{1}} - \ket{0_{0}} \right)
\, .
\eae
Integrating back \wrt{} $b$ and changing integration variable we
recover our earlier result~(\ref{f01}) (similarly for the
spin-down component).
\section*{Acknowledgments}
The first author (C.{} C.) would like to acknowledge partial support 
from CONACyT projects 32307-E, 41208-F, and DGAPA-UNAM
projects IN 119792, IN 114302.

\end{document}